\documentclass[%
superscriptaddress,
longbibliography,
twocolumn,
 amsmath,amssymb,
 aps,
]{revtex4-1}

\usepackage{siunitx}
\usepackage{graphicx}
\usepackage{dcolumn}
\usepackage{bm}
\usepackage{hyperref}
\usepackage{subcaption}
\usepackage{floatrow}

\usepackage{siunitx}
\usepackage{color}

\newcolumntype{C}[1]{>{\centering}m{#1}}

\newcommand{\eref}[1]{Eq.~(\ref{#1})}
\newcommand{\fref}[1]{Fig.~\ref{#1}}

\begin{document}


\title{Electrode configuration and  electrical dissipation of mechanical energy in quartz crystal resonators}


%
\author{A. J. V\"{a}limaa}
\affiliation{Department of Applied Physics, Aalto University, P.O. Box 15100, FI-00076 AALTO, Finland}%

\author{J. T. Santos}
\affiliation{Department of Applied Physics, Aalto University, P.O. Box 15100, FI-00076 AALTO, Finland}%

\author{C. F. Ockeloen-Korppi}
\affiliation{Department of Applied Physics, Aalto University, P.O. Box 15100, FI-00076 AALTO, Finland}%

\author{M. A. Sillanp\"a\"a}
\affiliation{Department of Applied Physics, Aalto University, P.O. Box 15100, FI-00076 AALTO, Finland}%

\begin{abstract}
\noindent Mechanical resonators made with monolithic piezoelectric quartz crystals are promising for studying new physical phenomena. High mechanical quality factors ($Q$) exhibited by the mm-sized quartz resonators make them ideal for studying weak couplings or long timescales in the quantum regime. However, energy losses through mechanical supports pose a serious limiting factor for obtaining high quality factors. Here we investigate how the $Q$ of  quartz resonators at deep cryogenic temperatures  can be limited by several types of losses related to anchoring. We first introduce means to reduce the mechanical losses by more than an order of magnitude in a no-clamping scheme, obtaining $Q$-factors of $10^8$ of the lowest shear mode. We can exclude a wide coverage of aluminum metallization on the disk or bond wires as sources of dissipation. However, we find a dramatic reduction of the $Q$-factor  accompanying an electrode configuration that involves strong focusing of the vibrations in the disk center. We propose a circuit model that accounts for the reduced mechanical $Q$-factor in terms of electrical losses.  In particular, we show how the limiting factor for losses can be small ohmic dissipation in a grounding connection, which can be interpreted as electrical anchor losses of the mechanical device.

\end{abstract}

\maketitle
\section{\label{sec:level1} Introduction}

Bulk Acoustic Wave (BAW) piezoelectric mechanical resonators are in use as the timebase in almost every digital electronics device \cite{Beek2011}. Besides oscillator use, the applications of BAW range  from scanning force microscopy \cite{Itoh1993, TANSOCK19921464}, filtering of analog circuits \cite{MMCE:MMCE20550} and magnetic-resonance force microscopy \cite{Mamin, Zhang}; and prospects in mechanical computation \cite{Masmanidis780}, mass sensors  \cite{Jensen} and quantum limited position measurement \cite{Cho36, Knobel}. Some more exotic possible applications include gravitational wave detection \cite{PhysRevD.90.102005}, quantum information manipulation \cite{PhysRevB.93.224518} and detection of specific types of dark matter \cite {PhysRevLett.116.031102}.

Thickness shear mode quartz resonators are a specific type of BAW where the piezoelectric crystal is usually an AT cut rectangular, square or disk plate with the displacement perpendicular to its thickness. The structure will resonate at frequencies for which the thickness is an odd multiple of half the acoustic wavelength. This type of resonator is extensively used as a sensor, for example for thin film deposition control, gas detection and liquid viscosity measurements. In these applications, it is often called the Quartz Crystal Microbalance (QCM), extensively studied after being proposed in Ref.~\cite{Sauerbrey1959}.

An important challenge with most of the thickness shear mode sensors is the difficulty of obtaining spatial sensing homogeneity. Usually the mechanical mode is not homogeneous across the sensing surface, so the sensitivity varies depending on the location of the load. This is particularly important in the CQM. For this reason there has been growing research interest in the optimization of the electrode geometry for uniform mass sensitivity. Some of the proposals include ring \cite{4716659} or elliptical \cite{1674-1056-20-4-047701} electrodes, or variable thickness metallization \cite{WANG2008150}.

Using plano-convex lens shaped quartz disks is a common practice that enhances the sensitivity uniformity by confining the mechanical mode to the center of the resonator plate, with the bonus of decreasing the clamping losses. In our earlier study \cite{1367-2630-19-10-103014} we introduced a design wit a set of large thin-film aluminum grounded spikes, extending from the edges to the center of the disk. The spikes were used to further focus the mechanical mode of a plano-convex crystal by enhancing the electrical field in the center. The focusing of the mode shape not only increases the spatial detection consistency but also the sensitivity in the smaller detection area, due to a bigger localized strain. The focusing can also be very useful for microfluidic / chemical / biological sensors, where reactive species are immobilized in a specific region of the substrate and the mode shape can be focused directly and exclusively below that active area.


The motivation of the work in Ref.~\cite{1367-2630-19-10-103014} was to operate a quartz disk resonator near the quantum regime of the mechanical vibrations of the lowest shear mode. In that design, a Cooper-pair transistor (SSET) \cite{PhysRevLett.78.4817, PhysRevLett.72.2458} is coupled in parallel to an LC tank circuit  \cite{PhysRevLett.93.066805}, forming an effective cavity optomechanical setup. The strength of the interaction between microwave "light" and mechanical motion depends on the fraction of total piezoelectric surface charge that couples to the island of the SSET. The charge focusing scheme improves the overlap between the mechanical mode and the small SSET island area of a few square micrometers, and enhances the interaction by two orders of magnitude.

\begin{figure*}
    \begin{subfigure}[b]{0.3\textwidth}
        \includegraphics[width=\columnwidth]{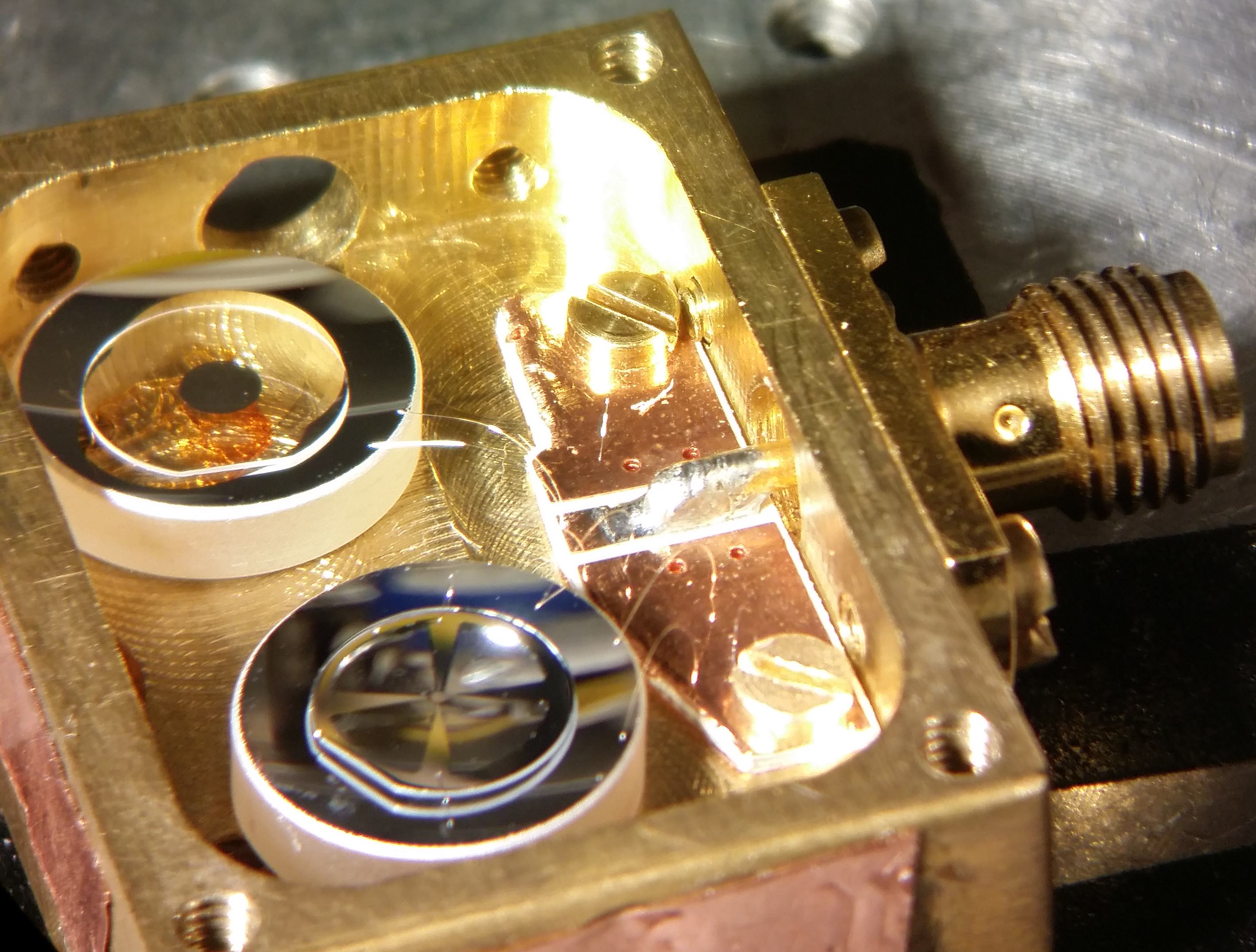}
 		\caption{} \label{fig:1boxphoto}
	\end{subfigure}
	\begin{subfigure}[b]{0.25\textwidth}
    	\centering
  		\includegraphics[width=.8\columnwidth]{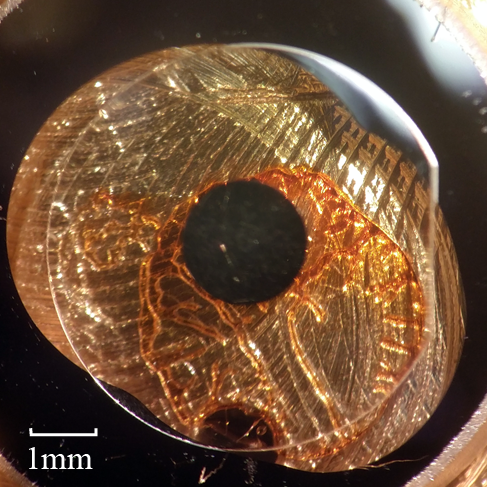}
  		\caption{} \label{fig:Al-i}
	\end{subfigure}
    \centering
    \begin{subfigure}[b]{0.35\textwidth}
  		\centering
        \includegraphics[width=\columnwidth]{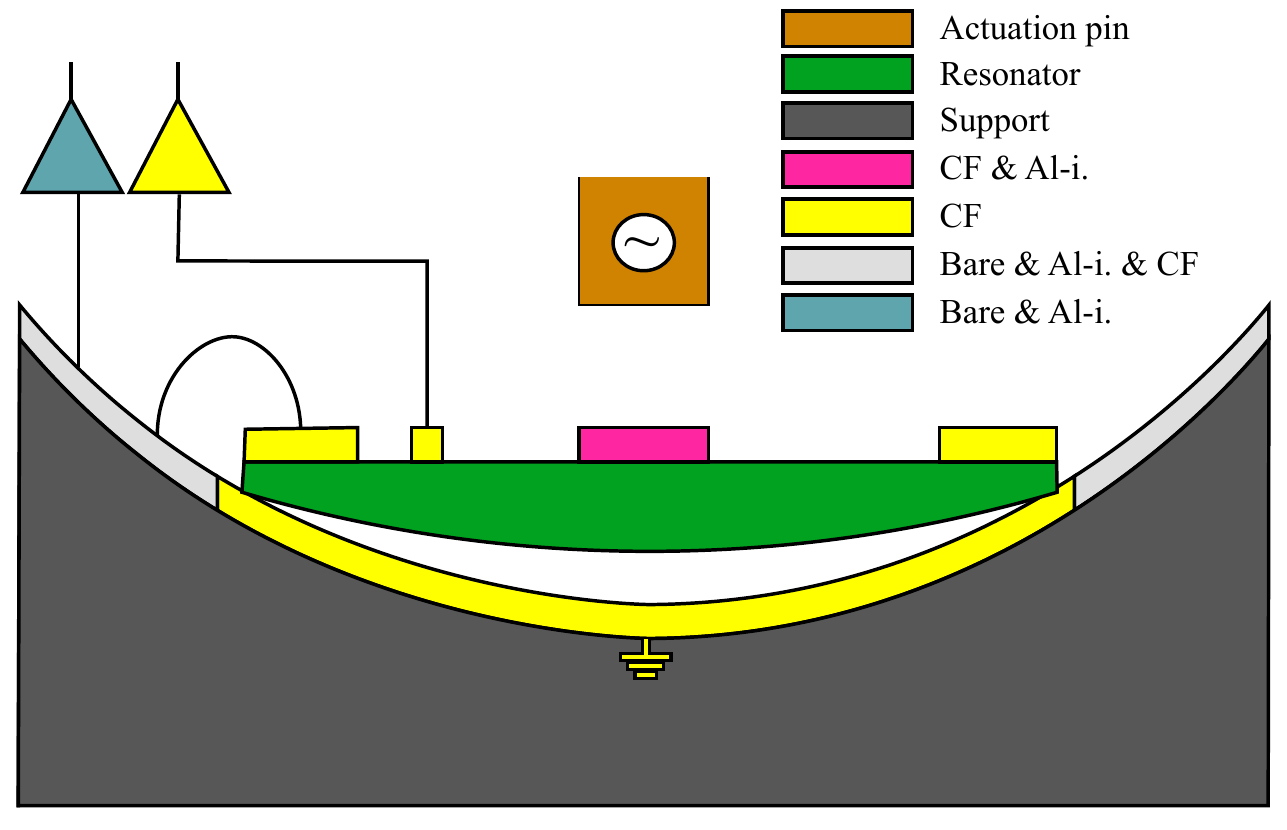}
 		\caption{} \label{fig:1side}
	\end{subfigure}
    \par\medskip
	\begin{subfigure}[b]{0.3\textwidth}
    	\centering
  		\includegraphics[width=.8\columnwidth]{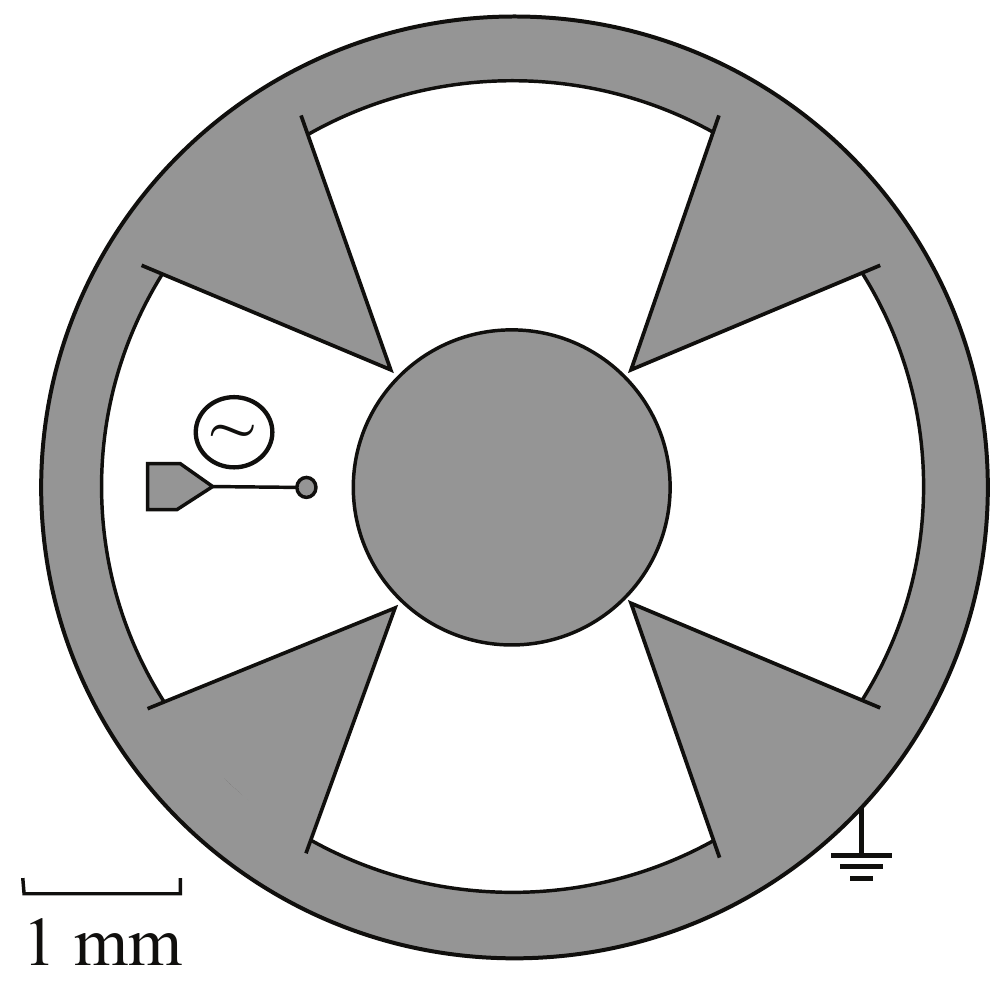}
  		\caption{} \label{fig:CF}
	\end{subfigure}
	\begin{subfigure}[b]{0.3\textwidth}
    	\centering
 		 \includegraphics[width=0.9\columnwidth]{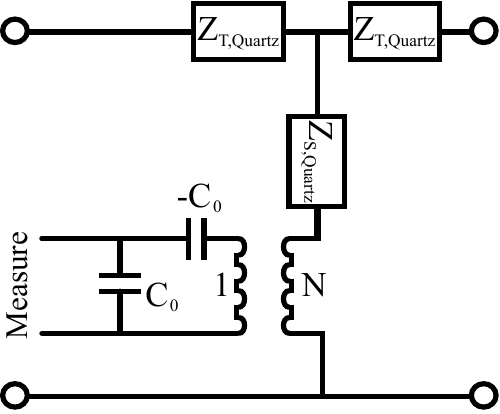}
 		 \caption{} \label{fig:1quartzmason}  
	\end{subfigure}
	\begin{subfigure}[b]{0.3\textwidth}
    	\centering
 		 \includegraphics[width=0.9\columnwidth]{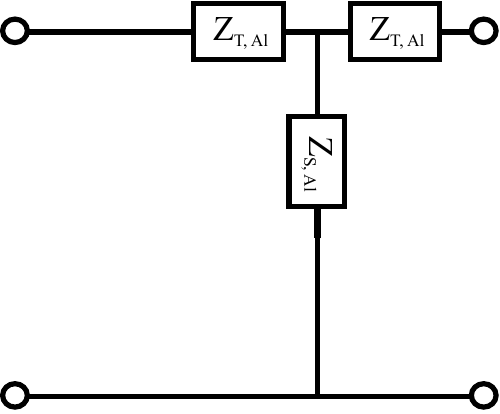}
 		 \caption{} \label{fig:1topmason}  
	\end{subfigure}
	\caption{a) Photograph of the actuation and transmission measurement setup employed to acquire the resonance frequency and quality factor of the quartz resonator at temperatures ranging from \SI{10}{\milli\kelvin} to \SI{300}{\kelvin}. Two quartz disk devices are seen laying in lens-shaped supports. b) Top-view photograph of a quartz disk having an aluminium island (black circle) in the center. c) Side view of the measurement schemes for the three different device schemes studied in this work: bare quartz (bare), aluminum island (Al-i.) and charge focusing (CF). The schemes are differentiated by colour coding. d) Charge focusing electrode configuration. e) Mason's model equivalent circuit for a piezoelectric layer. f) Mason's model for top / bottom coating layer. The drawn impedances are acoustic impedances.}
\end{figure*} 

In order to eliminate thermal noise in sensitive measurements, the quartz resonator needs to be cooled to the ground state of the mechanical vibrations. For the aforementioned resonator, the ground state corresponds to a very low temperature of about \SI{0.2}{\milli\kelvin}, as a consequence of its low resonance frequency $\approx \SI{7}{\mega\hertz}$. This temperature is well below the temperature achievable in a dilution cryostat, however optomechanical sideband cooling \cite{RevModPhys.86.1391}  may be used to further reduce the mechanical mode temperature below that of the environment. The cooling power of such a technique is linked to several factors, in particular the coupling strength in the effective optomechanical system, and the mechanical quality factor ($Q$). The  focusing scheme with grounded spikes enhances the coupling, however, we found that $Q \sim 10^6$, several orders of magnitude smaller than in principle possible in monolithic quartz resonators. One can thus suspect  that the focusing scheme adds mechanical energy leaking pathways, thus strongly deteriorating the prospects of sideband cooling performance.

In this work we investigate the mechanisms that cause the charge focusing setup to exhibit reduced $Q$, and how to minimize them. We present a model that includes such mechanisms, which can also be extrapolated to other electrode configurations. 
Similar to \cite{1367-2630-19-10-103014}, for the experimental work we will use \SI{6}{\milli\meter} diameter, \SIrange{200}{250}{\micro\meter} thick plano-convex quartz disks with a fundamental thickness shear mode around \SI{7}{\mega\hertz}. We experimentally measure the dissipation due to charge focusing and thin film electrodes from room temperature down to mK temperatures. In contrast to Ref.~\cite{coatinglosses}, we find that dissipation due to uniform thin film electrodes on the quartz surface is negligible, and instead the Q factor is limited by the electrical properties of the grounded charge focusing structures.


\begin{figure*}
	\begin{floatrow}
		\killfloatstyle
		\floatbox[\capbot]{table}[\FBwidth]{%
  		    \begin{tabular}{lc}\hline
        			Parameter & \\ 
       			 \hline
        			\rule{0pt}{4ex}$C_0$ & $\dfrac{\epsilon_{33} A_P}{t_P}$ \\
			\rule{0pt}{4ex}$C_{Spikes}$ & $\dfrac{\epsilon_{33} A_{Spikes}}{t_P}$ \\
       			 \rule{0pt}{4ex}$N$ & $C_0 h_{33}$\\
			 \rule{0pt}{4ex}$M$ & $C_{Spikes} h_{33}$\\
      			  \rule{0pt}{4ex}$Z_{0,n}$ & $A_n \sqrt{\rho_n c_{33,n}}$\\
     			   \rule{0pt}{4ex}$\Gamma_{n}$ & $\omega \sqrt{\dfrac{\rho_n} {c_{33,n}}}$\\ 
     			   \rule{0pt}{4ex}$Z_{T,n}$ & $i Z_{0,n} \tan\left(\dfrac{\Gamma t_n}{2}\right)$\\
     			   \rule{0pt}{4ex}$Z_{S,n}$ & $-i Z_{0,n} \csc\left(\Gamma t_n\right)$\\ \hline
     	   	    \end{tabular}
		}{%
  		\caption{Expressions for the parameters used in the Mason's and Modified Mason's models.}\label{tab:1}%
		}
		\killfloatstyle
		\ffigbox[\FBwidth]
		{\includegraphics[]{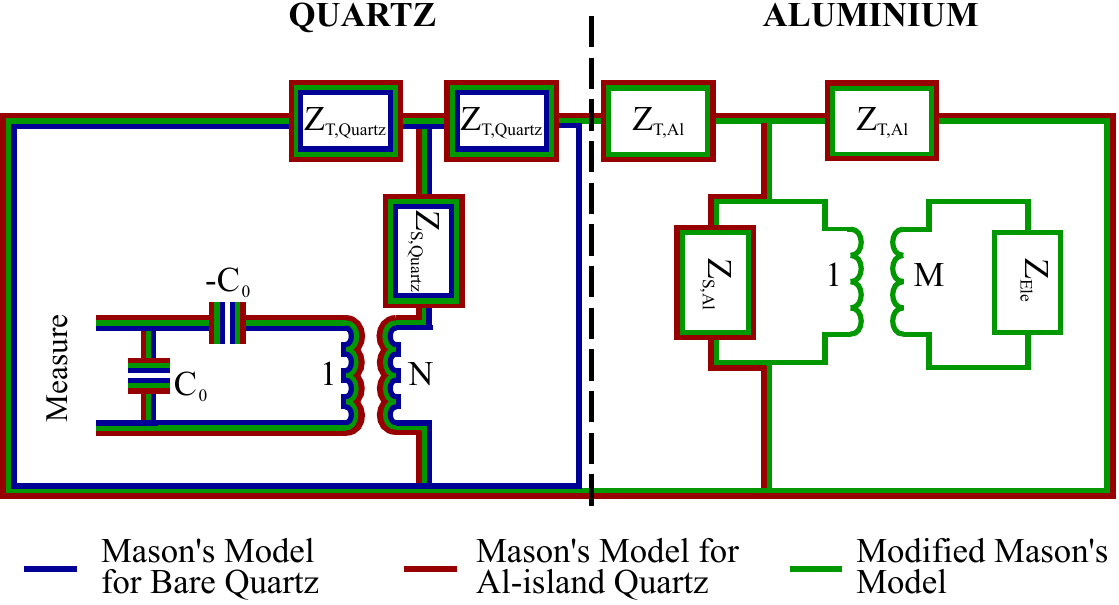}}
		{\caption{Schematic representation of the equivalent electrical circuits used  to model the shear thickness mode response of a quartz disk resonator with different electrode configurations. The blue and red traces represent the classical Mason's model for the situation of a bare quartz and Al-island quartz, respectively. The green trace represents the modified Mason's model proposed in this work to model mechanical energy leakage due to ohmic losses in the charge focusing electrode layout.} \label{fig:2}}
		\killfloatstyle
	\end{floatrow}
\end{figure*}

\section{\label{sec:level2} Mason's Model}

To study the dynamic behaviour of piezoelectric resonators, it is practical to use simple models expressed in a circuit representation. There are a number of different models available, from which some of the most common are the Butterworth-Van Dyke (BVD) \cite{Dye1925, Ballato}, Mason's \cite{mason1948electromechanical} and Krimholtz, Leedom and Matthae (KLM) \cite{4234776}. We use Mason's model, which compared to BVD has the advantage of clearly separating the electrical and acoustic domains and relating the material's physical properties to distinct acoustic impedances. 

The Mason's model for a piezoelectric layer is shown in Fig.~\ref{fig:1quartzmason} for the thickness mode. The piezo resonator is modeled as an acoustic transmission line coupled to its electrical counterpart by an ideal transformer with the number of turns $N$ proportional to the quartz electromechanical coupling coefficient. 
Notice that $N$ is a dimensional quantity, representing the transduction between acoustic and electrical impedances.
In the case of bare quartz, with no extra layers deposited on surfaces, the acoustic ports are shorted as depicted by the blue trace of Fig.~\ref{fig:2}. If another material is added on any of the surfaces, a  module similar to the one shown in Fig.~\ref{fig:1topmason}  is added to the corresponding port on the Mason's piezoelectric representation. The red trace of Fig.~\ref{fig:2} shows such representation for the quartz disk with the top surface coated. In our experiment, the coatings are the electrodes made out of aluminium.

The relationships between the materials constants and the model parameters are shown in Table~\ref{tab:1} \cite{849139}, where $\epsilon_{33}$ and $h_{33}$ are the piezoelectric material's permittivity and deformation factor, $c_{33,n}$ and $\rho_n$ the elastic stiffness and density of the $n$ (quartz or Al)-layer material, $A_P$ and $t_P$ the piezoelectric disk surface area and thickness, $A_n$ the $n$'s layer surface area, and $Z_{0,n}$ represents the specific acoustic impedance of material $n$. It is important to notice that in order to take into account the different sources of intrinsic dissipation (dielectric, elastic, ...) $\epsilon_{33}$, $h_{33}$ and $c_{33,n}$ are defined as complex numbers, with the imaginary part being the corresponding loss tangent.

The impedance $Z_{S,n}$ on the acoustic transmission line portrays the bulk elastic response of the $n$ material to the acoustic wave, while $Z_{T,n}$ mimics the behavior of such wave at the interfaces with adjacent material layers. In the case under study, namely a quartz disk with Al metallization on the top surface, the impedance matching between $Z_{T}$ of the quartz layer and the coating film total equivalent impedance will define the effect of the film on the resonator. Impedance mismatch creates wave reflections at the quartz / Al interface that can interfere constructively or destructively, depending on the phase shift and attenuation suffered by the acoustic wave while traveling through the different materials.

The analysis of the standard Mason's representation of Fig.~\ref{fig:2} for a quartz disk coated with aluminium shows that the total acoustic impedance in such case is 
$Z_A=Z_{S,Quartz}+Z_{T,Quartz}\parallel \left(Z_{T,Quartz}+Z_{Al}\right)$, where $Z_{Al}$ is the total acoustic impedance of the aluminium layer given by $Z_{Al}=Z_{T,Al}+Z_{T,Al}\parallel Z_{S,Al}=i\rho_{Al} A_{Al} \nu_{Al}\tan\left(\omega t_{Al}/\nu_{Al}\right) $ and $\nu_{Al}$ is the complex acoustic velocity in the aluminium. Depending on the thickness of the Al film, its effect on the piezoelectric resonance can be described either as mass damping in the regime with $t_{P}\gg t_{Al}$, or as radiation damping in the regime $t_{P}\ll t_{Al}$.

In the mass damping regime with a thin film coating, $\tan\left(\omega t_{Al}/\nu_{Al}\right)\approx \omega t_{Al}/\nu_{Al}$, which yields $Z_{Al}=i m_{Al} \omega$. In this case the aluminium mass acts as an inductance on the total acoustic impedance of quartz, so the resonance peak is shifted but the effect on the mechanical quality factor is negligible. This means loading of the mechanical resonator by ideal mass, with the film moving synchronously with the quartz. In our experiment the quartz thickness $t_P \approx \SI{200}{\micro\meter} \, ... \, \SI{250}{\micro\meter}$ and the aluminium thin film thickness $t_{Al} \approx \SI{30}{\nano\meter}$, and hence we expect to be deep in the mass damping regime.

\begin{figure*}[htp]
  \ffigbox[\FBwidth]
  {\begin{subfloatrow}
      \ffigbox
        {\caption{} \label{fig:3a}}%
        {\includegraphics[scale = .5]{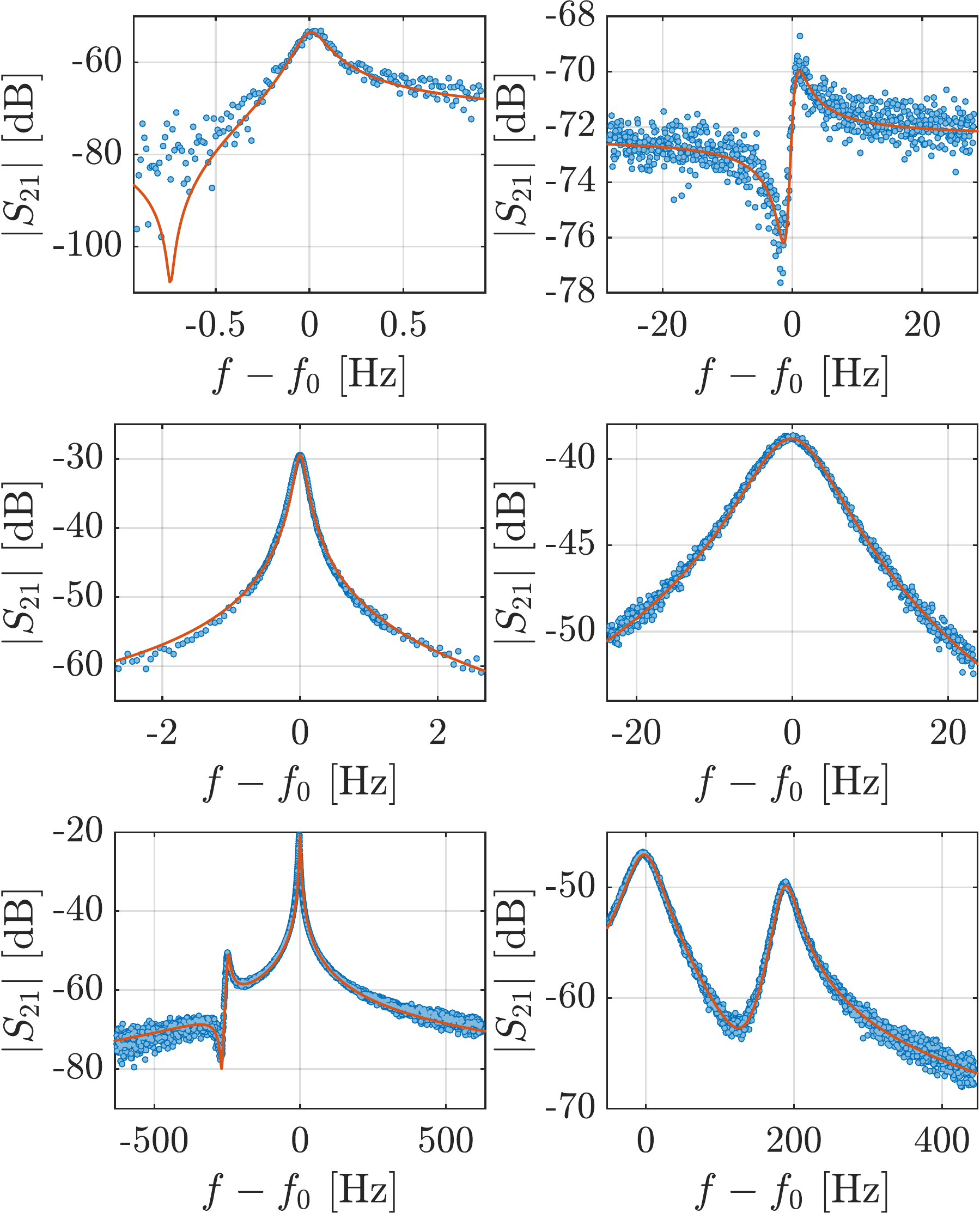}}
        \ffigbox
        {\caption{}\label{fig:3b}}%
        {\centering\includegraphics[]{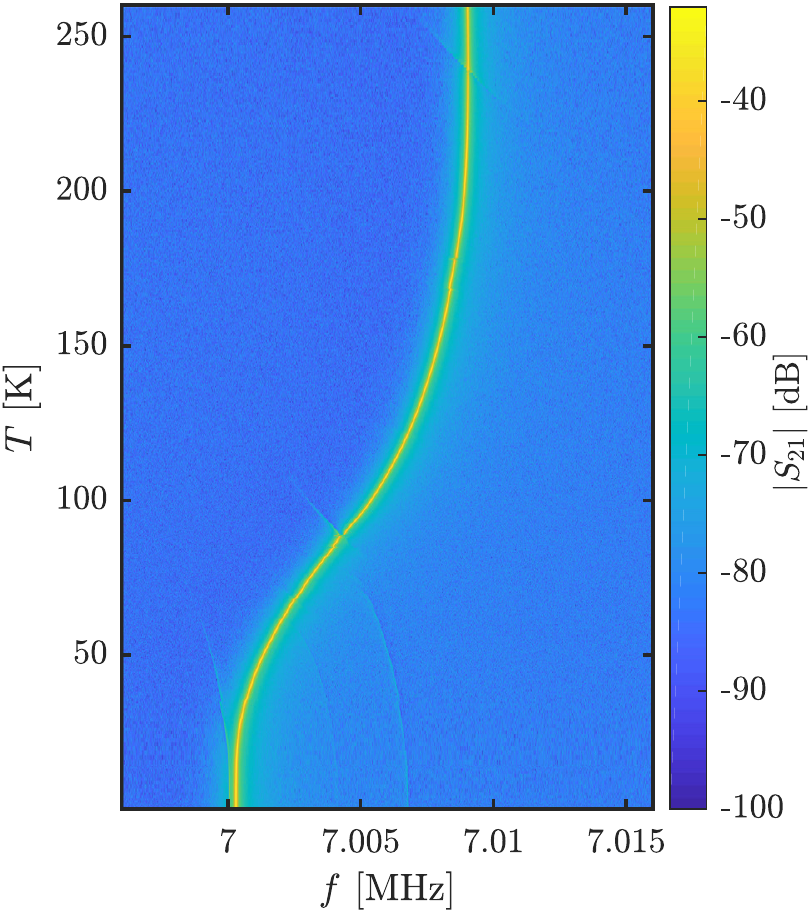}\vskip0.9cm}
  \end{subfloatrow}}
  {\caption{\subref{fig:3a}) The magnitude characteristics of the transmission coefficient of the mechanical mode is illustrated at the base temperature (left column) and above \SI{5}{\kelvin} (right column) for: bare quartz (top row), Al coated quartz (center row) and one of the avoided crossing measured for the charge focusing electrode configuration (bottom row). The measurement data (blue dots) is fitted in the complex plane, but projected on the magnitude plot (solid red line). The strong attenuation in the lines is compensated by a LNA that provides an amplification of 56 dB. \subref{fig:3b}) Thermal drift of the mechanical resonance peak for the CF electrode configuration caused by the change of the quartz material coefficients as a function of temperature. It is possible to observe several other weakly coupled modes to cross (in fact, exhibiting small avoided crossings) the main peak at temperatures above \SI{40}{\kelvin}, which, as observed in Fig \ref{fig:4} and \ref{fig:5}, impact negatively the quality of the acquired data above such temperature.}}
\end{figure*}



\section{Experimental techniques}

The lowest shear mode at $\omega_0/2\pi \approx \SI{7}{\mega\hertz}$ is characterised in the transmission scheme of \fref{fig:1side}. The transmission setup allows fast and accurate determination of the resonance peak even when the mode frequency is changing with temperature, and the system is only weakly loaded by the measurement ports.

The quartz disk resonator lies unanchored on top of a concave quartz lens with the convex part on the bottom, maintaining the central region of the disk suspended. In the Al-coated quartz electrode configuration of \fref{fig:Al-i}, there is an Al-island of 2 mm diameter and $\SI{30}{\nano\meter}$ thickness evaporated on the top surface of the piezo disk. In this configuration there are no bonding wires connected to the resonator, but it is actuated and measured through a top port (SMA pin of 1 mm radius at roughly 200 $\mu$m distance) and bottom electrode, as seen in \fref{fig:1side}. The bottom electrode lies on the concave support chip, that is not fully metal-covered to control the external coupling through the measurement port. The same method of actuation is used for the the bare quartz sample.

The charge focusing (CF) design requires grounding of the focusing spikes (Figs.~\ref{fig:1side} and \ref{fig:CF}) and below the resonator for proper functioning. Hence, the supporting lens structure is fully metalized, grounded, and wire-bonded to the quartz disk. In separate measurements, we have verified that the bond wires near the disk edges do not affect the quality factor of the resonator. 

The transmission measurement setup involves the ports on top of the resonator and on-chip (charge focusing scheme) or in the bottom chip (for bare and Al-coated quartz), as depicted in  \fref{fig:1side}. The resonance peak is tracked from room temperature to base temperature of the refrigerator. The tracking algorithm needs to account for a drift of 10 kHz in the resonance frequency, that is roughly $10^5$ times the linewidth. 
The  resonance line profiles are fitted in the complex plane by a model  that is derived from a 2-port RLC-circuit.
The model is also used to evaluate the external couplings of the ports from the reflection parameters, allowing to separate the total quality factor ($Q_{Tot}$) into internal ($Q_{int}$), and the external quality factor ($Q_{ext}$) set by the measurement ports. We note that apart from the losses due to the measurement ports, $Q_{int}$ includes all losses, also the electrical losses in the charge focusing scheme as discussed below. Our model also  allows us to consider several overlapping modes.
%
%
The characteristics of the transmission peak data and the fitting are illustrated in \fref{fig:3a} for the three configurations at the base temperature ($\approx \SI{10}{\milli\kelvin}$) and at high temperatures above \SI{5}{\kelvin}. 

\begin{figure*}
   		\includegraphics[]{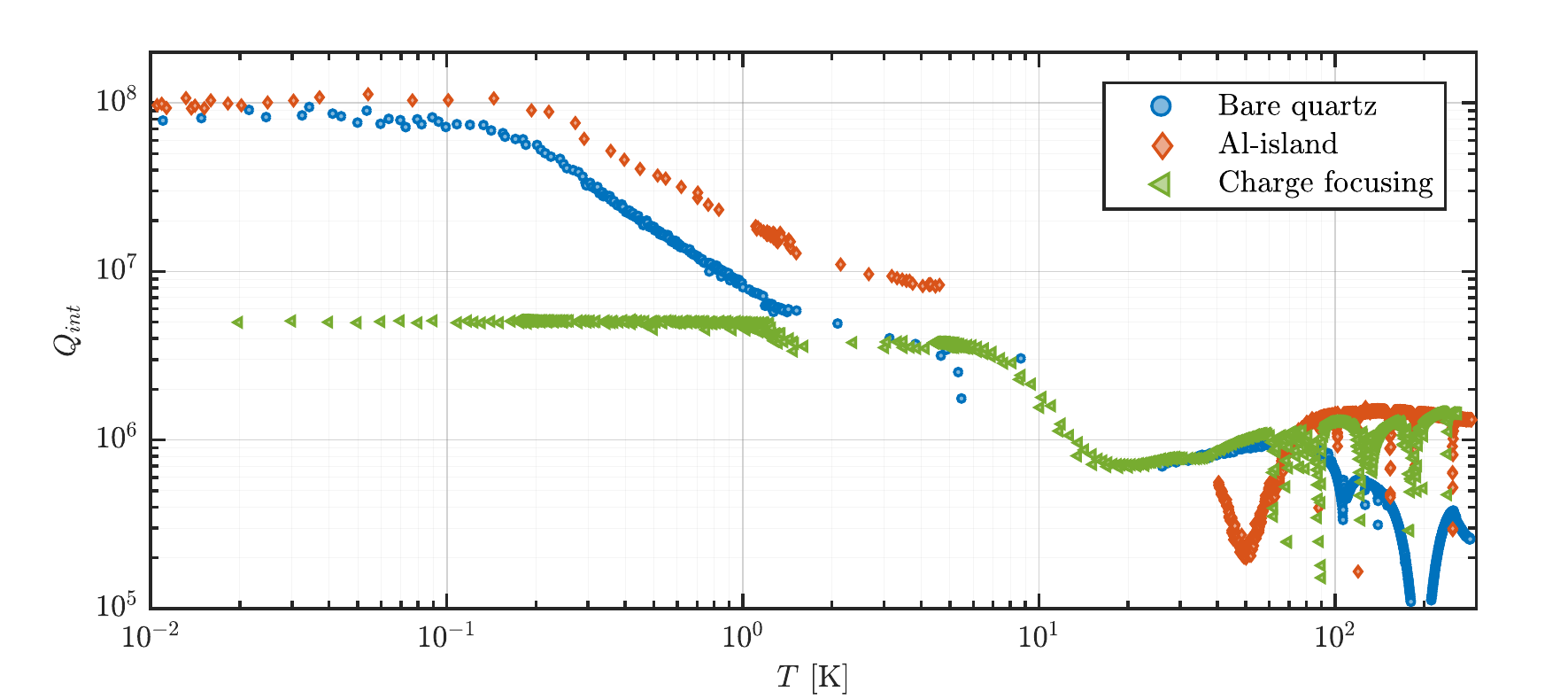}
  		\caption{ Dependence of the internal mechanical quality factor on temperature for the three metallization configurations studied in this work: bare quartz disk (blue dots), quartz disk coated with an aluminium thin film island (red diamonds), and quartz disk  coated with an aluminium thin film in a charge focusing configuration (green triangles).} \label{fig:4}
\end{figure*}

Figure \ref{fig:4} shows the measured quality factor of the three aluminium metallization configurations under study. 
One can see that, as expected from the minimal effect of the thin Al layer, the quality factors for bare quartz and the Al-island quartz exhibit similar temperature dependence. However, the charge focusing design has a much lower $Q$ than the other two configurations at temperatures below $\sim 10$ K, while at higher temperatures all the configurations show roughly similar $Q$ values. The sharp dips in $Q$ above $\sim 40$ K result from crossing of the main mode with other weakly coupled modes that have an opposite temperature dependence (\fref{fig:3b}) of the resonance frequency. The fact that the crossing spurious modes possess lower $Q$ factors, is reflected in the broader linewidths of the hybridised modes. Examples of split peaks are shown in \fref{fig:3a} with a coupling in the order of 200 Hz.

The charge focusing case will be discussed below, and now we focus on the bare quartz disk and Al-island quartz disk configurations. The measured internal quality factors of the bare and Al-island coated quartz at the base temperature are approximately 80 million and 100 million, respectively. The Al-island quartz, somewhat surprisingly, exhibits the best $Q$. We believe this is due to small variations from chip to chip in some dissipation channels, which becomes more evident at lower temperatures when other dissipation decreases. Below 100 mK the $Q$ factor saturates, which can be either due to insufficient thermalisation of the driven resonator, or a temperature-independent dissipation source.

%

We use the material parameters of Table \ref{tab:2} and apply them to the calculation of the Mason's equivalent circuit parameters of Table \ref{tab:1}. The literature usually provides the material coefficients only for room temperature. To check the model at low temperature we fit the measurements of the bare quartz at base temperature to the circuit represented by a blue trace in  Fig.~\ref{fig:2} to find the loss tangents.  Then we use the fitted quartz complex coefficients to calculate the quality factor expected for the mass-loaded Al-island quartz disk  (circuit of Fig.~\ref{fig:2}, red trace). The quartz coefficients fitted at the base temperature show the loss tangents for the complex permittivity and stiffness approximately two orders of magnitude lower than the literature values at room temperature. This is qualitatively reasonable, because at lower temperatures the dielectric and elastic losses are expected to be smaller.

\begin{table}\begin{tabular}{ccc}\hline
        			  & Literature Value \cite{sherit2002} & Fitted Value \\ 
			  & (Room Temperature) & (Base Temperature) \\ 
       			 \hline
        			\rule{0pt}{4ex}\begin{tabular}{c}$\epsilon_{33}$\\(\SI{e-11}F/m)
\end{tabular}&$\num{5.3}\cdot (1-\num{2E-3}i)$&$\num{5.3}\cdot(1-\num{4.0E-5}i)$ \\
			\rule{0pt}{4ex}\begin{tabular}{c}$c_{33,Quartz}$\\(\SI{e10}N/m\textsuperscript{2})
\end{tabular}&$\num{2.94}\cdot(1+\num{4.5E-6}i)$&$\num{2.94}\cdot(1+\num{8.0E-8}i)$ \\
			\rule{0pt}{4ex}\begin{tabular}{c}$h_{33}$\\(\SI{e9}V/m)
\end{tabular}&$\num{1.67}\cdot(1+\num{2.9E-5}i)$&$\num{1.67}\cdot(1+\num{2.9E-5}i)$ \\
			\rule{0pt}{4ex}\begin{tabular}{c}$\rho_{Quartz}$\\(Kg/m\textsuperscript{3})
\end{tabular}&$\num{2650}$&$\num{2650}$ \\
			\rule{0pt}{4ex}\begin{tabular}{c}$c_{Al}$\\(\SI{e10}N/m\textsuperscript{2})
\end{tabular}&$\num{8}\cdot(1+\num{1E-2}i)$&$\num{8}\cdot(1+\num{1E-2}i)$ \\
			\rule{0pt}{4ex}\begin{tabular}{c}$\rho_{Al}$\\(Kg/m\textsuperscript{3})
\end{tabular}&$\num{2700}$&$\num{2700}$ \\
\hline
\end{tabular}
\caption{Complex material coefficients of AT cut quartz and aluminium at room temperature and the values obtained from the fit of the Mason's equivalent circuit (Fig. \ref{fig:2}, blue trace) to the experimental results shown on Fig. \ref{fig:4} for bare quartz at base temperature ($T\approx\SI{10}{\milli\kelvin}$).}\label{tab:2}
\end{table}

We thus find that the 30 nm thick aluminium layer is thin enough to keep the loading of the quartz resonator close to the ideal mass damping region, where the surface coating losses are small or negligible. Since the charge focusing design of Fig.~\ref{fig:CF} has roughly similar amount of metalization, the lower quality factors observed with charge focusing cannot result from coating losses in the electrode metalization, in contrast to the findings in Ref.~\cite{coatinglosses}.


\section{\label{sec:level3} Modified Mason's Model}

We continue the discussion on the  mechanical $Q$ values of the charge focusing electrode configuration (\fref{fig:CF}). As shown in  Fig.~\ref{fig:4}, the internal $Q$ for the  focusing configuration below $\sim 100$ mK is more than an order of magnitude smaller than in the other two setups and constant below $T\approx \SI{1}{\kelvin}$. 
At $T \sim 1.2$ K, we observe a sharp step down, followed by a plateau and then fast drop up to $T\approx \SI{20}{\kelvin}$. At higher temperatures $Q$ increases again and settles down to values similar to the other two configurations. The sharp step down of $Q$ at $T \sim 1.2$ K  matches the superconducting critical temperature of aluminium, best seen in Fig.~\ref{fig:5}, an indication that the low $Q$ of the focusing design may be connected to the conductivity of the aluminium thin-film coating. 

Next we discuss in detail the hypothesis that the resistance in the Al coating, or generally in the electrical part of the circuit, can be the source of low mechanical $Q$ factors. The strain-induced piezoelectric surface charge density in the vicinity of the Al coating capacitively couples to it, creating a charge distribution on the aluminium layer. If such layer is connected to the ground, for instance with wire bonds as in our case, the presence of any electrical resistance between the coupled charges and the ground creates a potential difference across the aluminium thin-film. The potential difference will drive the charges to the ground electrode, dissipating energy in the process. If the resistance to ground is too big (eg: coating layer not connected to ground; non-conductive coating material) the charge current is hindered and no dissipation occurs. Similarly, zero resistance implies no losses, and in between, there is expected to be a regime of rough impedance match and maximum losses.


The leakage of mechanical energy to the electrical domain through the coating layer is not accounted for by the classical Mason's model. We propose a modification to the model in order to be able to describe the low $Q$ factors measured in the charge focusing configuration. In a way similar to the Mason's representation of a coupling between the quartz vibrations and the external measurement circuitry (Fig~\ref{fig:1quartzmason}), we include an ideal transformer to represent the coupling to the electric domain in the aluminium layer. As with the measurement circuit, the transformer winding ratio is proportional to the quartz electromechanical coupling and, in the present case, the area of the grounded spikes. The modified Mason's model is shown by the green trace of Fig.~\ref{fig:2} and the winding ratio $M$ can be calculated with the expression from Table \ref{tab:1}.

\subsection{\label{sec:level4}Temperature dependence of the electrical resistivity of aluminium}

We turn the discussion to estimation of the electrical impedance $Z_{Ele}$  (see \fref{fig:2}) of the circuit including the charge focusing spikes and ground connection. We start by assuming there is a temperature-independent resistance $R_{0}$ in the ground connection.
We also assume that the Al is pure enough to have a sharp superconducting transition. For temperatures above the aluminium superconducting critical temperature, $T_{C, Al}=\SI{1.2}{\kelvin}$, there is also a contribution from the aluminium film resistivity $\rho_{ele, Al}$. Hence, $Z_{Ele}(T)$ is given by:
\begin{equation}\label{eq:1}
Z_{Ele}(T)=
\begin{cases}
R_0\quad\quad\quad\quad\quad\quad\quad,\quad T<T_{C, Al}\\
R_0+G\rho_{ele, Al}(T)\quad,\quad T>T_{C, Al}\\
\end{cases}
\end{equation}
where $G$ is a geometric factor converting the resistivity of the aluminium coating to the average resistance experienced by the charges traveling through the spikes.

To calculate the aluminium film resistivity above its superconducting critical temperature we use Matthiessen rule:
\begin{equation}\label{eq:2}
\rho_{ele, Al}(T)=
\begin{cases}
0\quad\quad\quad\quad\quad\quad,\quad T<T_{C, Al}\\
\rho_{R}+\rho_{Ph}(T)\quad,\quad T>T_{C, Al}\\
\end{cases}
\end{equation}
where $\rho_R$ is a temperature-independent term arising from the electron scattering on the material's defects and impurities, and $\rho_{Ph}$ is a temperature-dependent term due to electron-phonon interactions.

Some studies have found that the main electron-phonon scattering mechanisms in nanostructured polycrystalline thin films are the scattering from intragranular atoms or atoms at grain boundaries, or at surfaces \cite{TOMCHUK, Hodak, Ma}. Both mechanisms are inversely proportional to the grain size of the material, which in thin films is approximately equal to the film thickness. As a consequence, our Al layer has a much higher resistivity than expected from bulk Al. According to Ref.~\cite{LijunSun} the contribution of the electron-phonon interaction to the resistivity of the thin film of Al can be calculated from
\begin{equation} \label{eq:3}
\rho_{Ph}(T)=a T^m + b T^n \,,
\end{equation}
where the values for the parameters $a$, $b$, $n$ and $m$ are given in Table \ref{tab:3}. Here, $a T^m$ denotes the resistivity contributed by intragranular atoms and $b T^n$ represents the resistivity related to grain boundaries or surfaces.

\begin{table}\begin{tabular}{cc}\hline
        	& Value \\        			 \hline
        	\rule{0pt}{4ex}$\quad\rho_{R}\quad$&$\quad\SI{6.69e-8}{\ohm\meter}\quad$ \\
            \rule{0pt}{4ex}$\quad a\quad$&$\quad\num{2.97e-18}\quad$ \\
            \rule{0pt}{4ex}$\quad m\quad$&$\quad\num{4}\quad$ \\
            \rule{0pt}{4ex}$\quad b\quad$&$\quad\num{6.39e-15}\quad$ \\
            \rule{0pt}{4ex}$\quad n\quad$&$\quad\num{3}\quad$ \\
\hline
\end{tabular}
\caption{Aluminium parameters for the calculation of the resistivity at different temperatures, see \eref{eq:3}. Values from \cite{LijunSun}.}\label{tab:3}
\end{table}

\subsection{\label{sec:level5}Mechanical quality factor of the charge focusing electrode configuration}

The electrical impedance $Z_{Ele}$ to ground experienced by the charges can be estimated for the case of the Al thin-film using Eqs.~(\ref{eq:1}), (\ref{eq:2}), (\ref{eq:3}) and the values given in Table \ref{tab:3}. Two of the parameters are used to fit the model to the experimental data: The $\SI{0}{\kelvin}$ residual resistance $R_0$, and the geometric conversion factor $G$ in \eref{eq:3}.
Figure \ref{fig:4} shows that the temperature dependence of the mechanical $Q$ in the charge focusing electrode configuration has a plateau when the aluminium is in the superconductive state, immediately followed by a sharp drop at $T>\SI{1.2}{\kelvin}$. Those two features are used to fit the two free parameters $R_0$ and $G$.

When aluminium is in the superconducting state $Z_{Ele} (T<\SI{1.2}{\kelvin})\approx R_0$, and $R_0$ can be estimated by finding how large $Z_{Ele}$ is needed in the modified Mason's model to obtain the plateau at $T<\SI{1.2}{\kelvin}$. We obtain that $R_0\approx \SI{1}{\ohm}$ and, as previously mentioned, it likely arises from resistance sources like the small contact points in wire bonding, oxide formation at the aluminium - wire bond interface. 
The parameter $G$ can be estimated from the step height observed at \SI{1.2}{\kelvin}. At this temperature $\rho_{Ph}\approx 0$; hence from Eqs.~(\ref{eq:1}) and (\ref{eq:2}) we obtain $\rho_{ele,Al}=\rho_R$ and $G=\left(Z_{Ele}(\SI{1.2}{K})-R_0\right)/\rho_{R}$. To find $Z_{Ele}(\SI{1.2}{K})$ we calculate the impedance needed in the modified Mason's model to reproduce the $Q$ factor measured in the plateau above $T>\SI{1.2}{\kelvin}$. The value of $G$ obtained from the fitting is $\approx \SI{6e6}{\per\meter}$.

In the  focusing  configuration (see Fig.~\ref{fig:CF}) the mechanical mode is mainly focused in the center of the disk, thus most of the piezo charges that couple to the aluminium will arise at the tip of the spikes, and travel $L_{Spikes}\approx \SI{3}{\milli\meter}$ to the grounded bond wires. The average cross section area of the metallization experienced by the charges traveling along the spikes is $W_e\times t_{AL}\approx \SI{1}{\milli\meter} \times t_{Al}$. We can hence roughly estimate the expected value of $G\sim L_e/(t_{Al} W_e)=\SI{e8}{\per\meter}$. The difference to the value obtained above indicates that a considerable number of electrons still couple to other regions of the spikes other than the tips and travel shorter distances to ground.

\begin{figure*}
  		\includegraphics[]{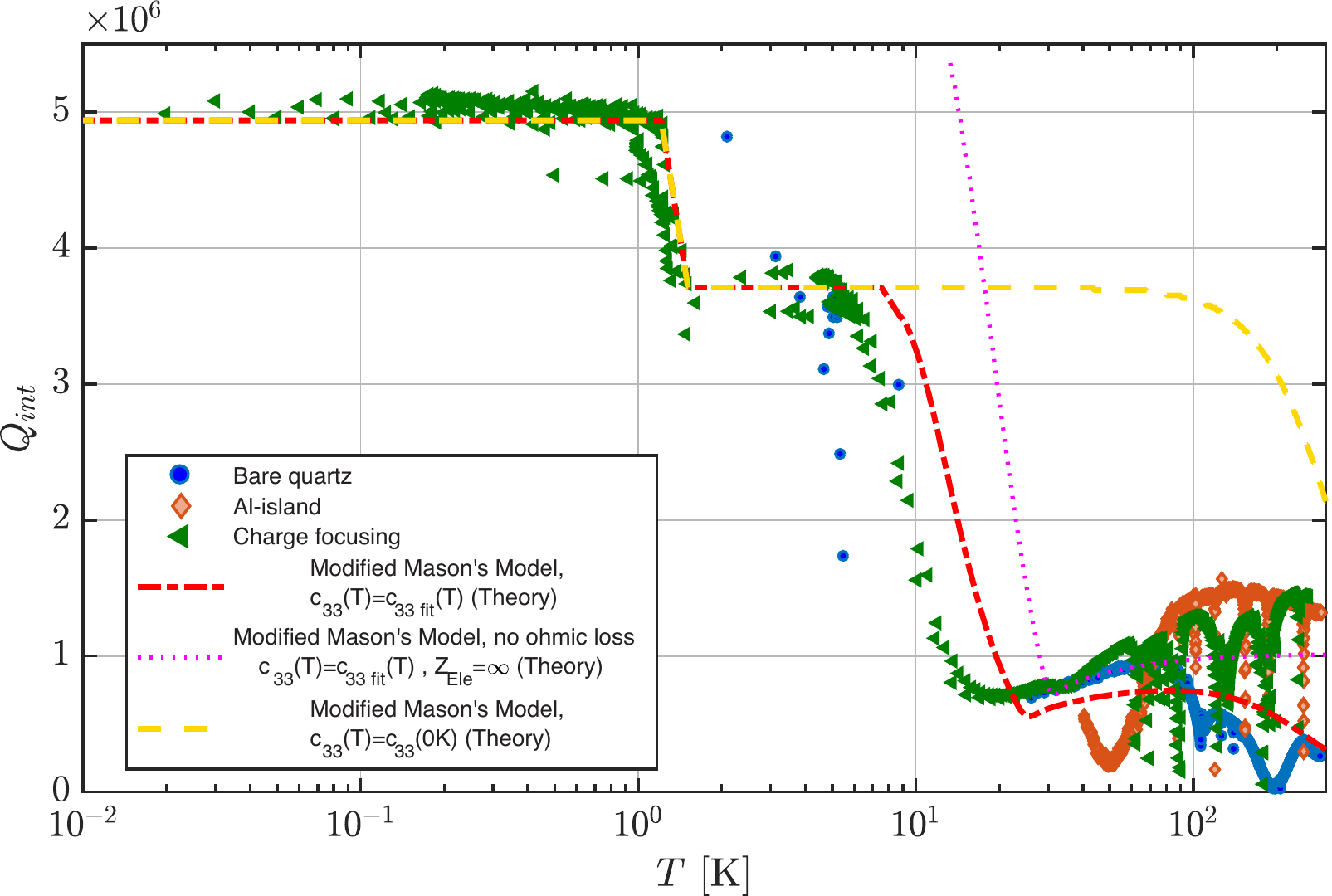}
  		\caption{Measurement of the temperature dependence of the total mechanical quality factor for the bare quartz and charge focusing electrode configurations. The lines display the theoretical curves obtained by using the modified Mason's model of Fig. \ref{fig:2} while considering: (dashed) $c_{33}$ of quartz constant in temperature and equal to the value fitted for the base temperature shown in Table \ref{tab:2}; (dotted) $c_{33}$ of quartz temperature-dependent, estimated for each $T$ from the fitting of the classic Mason's model to the bare quartz data of Fig. \ref{fig:4}, and $Z_{Ele}=\infty$, what reduces the modified model to the classic Mason's model; (dash-dot) $c_{33}$ of quartz temperature-dependent and estimated for each $T$ from the fitting of the classic Mason's model to the bare quartz data of Fig. \ref{fig:4}.} \label{fig:5}
\end{figure*}

Figure \ref{fig:5} shows a detailed view of the data in \ref{fig:4} for the charge focusing electrode configuration. The data is overlaid with three different theoretical curves: in the dashed curve the variation of the quartz intrinsic losses is neglected and the ohmic losses of aluminium are taken into account with the modified Mason's model; in the dotted curve the intrinsic quartz losses are taken into account and the ohmic losses are neglected by considering $Z_{Ele}=\infty$
; in the dash-dotted curve both intrinsic quartz losses and ohmic losses are taken into account using the modified Mason's model. We will next discuss in detail the three theoretical curves.

For the computation of the dashed curve the quartz material coefficients were considered constant in temperature, so the only source of temperature-dependence on the resonator dissipation comes from the aluminium spikes' ohmic losses. The drop of $Q$ with the increase of temperature for the bare quartz design, seen in Fig.~\ref{fig:4}, however, demonstrates that the intrinsic dissipation of quartz is not constant in temperature, explaining the discrepancy between the dashed line and the experimental data of Fig.~\ref{fig:5}.

The temperature dependence of the quartz material coefficients needs to be taken into account for properly modeling the system. For that purpose, we assume that the only coefficient changing with temperature is $c_{33}$ (specifically the imaginary part of $c_{33}$, the corresponding loss tangent) and calculate it for each temperature through the fit of the classic Mason's model to the experimental data for bare quartz and Al coated quartz displayed in Fig.~\ref{fig:4}. To minimize the aforementioned problems with variability of $c_{33}$ from chip to chip, the value used for subsequent data analysis is the average for each temperature of the values calculated from the two fits.

The bare quartz measurement data is difficult to interpret above $T\approx \SI{100}{\kelvin}$ because of several avoided crossings, clearly seen in  \fref{fig:3b}. The $c_{33}$ for those temperatures was extrapolated from the fit of a curve to the temperature range between \SI{20}{\kelvin} and \SI{80}{\kelvin}. The dotted line of Fig.~\ref{fig:5} represents the modified Mason's model calculated using the fitted values of $c_{33}$ and disregarding the ohmic losses by considering $Z_{Ele}=\infty$, effectively reducing the modified Mason's model to the Al coated classic Mason's model. 

The dash-dotted line
takes into account the temperature dependence of both the ohmic and quartz intrinsic losses as discussed.
This model provides the best match to the experimental data for the charge focusing setup, showing that in such electrode configuration both dissipation mechanisms (quartz's intrinsic mechanical dissipation and electrical ohmic dissipation) play a role in determining the total dissipation of the mechanical resonator. The slight difference between the dash-dotted curve and the experimental results at $\approx\SI{10}{\kelvin}$ is likely associated to uncertainty in the definition of the material coefficients for each individual chip. 
At very low temperatures, below \SI{6}{\kelvin}, the quartz intrinsic loss is negligible and the total dissipation is dominated by ohmic losses.

The  impedance of the spikes at room temperature 
is $Z_{Ele} (\SI{300}{\kelvin})\approx\SI{3}{\ohm}$. If instead of aluminium we would use a more resistive coating material, the model predicts that when $Z_{Ele}\approx\SI{20}{\ohm}$ there is impedance matching between $N^2 Z_{Ele}$ and $Z_{T,Al}$. For impedances above \SI{20}{\ohm} the current will flow preferentially through the $Z_{T,Al}$ branch of the circuit of Fig. \ref{fig:2} and the impact of the spikes electrical impedance on the $Q$ factor will decrease. For $Z_{Ele}$ much larger than the matching value, the ideal transformer that couples the electrical domain of the aluminium layer to the quartz acoustic mode will act almost as an open line, making our circuit similar to the standard Mason's scheme. 
Materials with higher resistivity than aluminium may mitigate the effect of mechanical energy leakage to the electric domain.

\section{\label{sec:level6} Conclusions}



In this work we studied the quality factor of monolithic quartz disk piezoelectric resonators with different electrode configurations, and at temperatures ranging from deep cryogenics (\SI{10}{\milli\kelvin}) up to room temperature. We found that even with substantial electrode coverage, it is possible to achieve quality factors up to $\sim 10^8$. The extra dissipation introduced by mechanical losses in thin-film of aluminium coating is negligible, a fact that can be explained by the small thickness of the aluminium layer (\SI{30}{\nano\meter})  compared to the quartz disk (\SI{200}{\micro\meter} to \SI{250}{\micro\meter}).

The use of a charge focusing design for the electrodes, where sharp grounded aluminium  spikes extend from the edges of the disk to its center, has an effect to drastically decrease the mechanical quality factor down to around \num{e6}. 
We suggest a model where the surface piezoelectric charge density can couple capacitively to the  spikes. The charges arising at the spikes will create a current to ground, dissipating energy in the non-negligible resistances at the aluminium thin-film.
This dissipation mechanism represents a leakage of mechanical energy to the electrical domain.

We  suggest a modification to the Mason's model that takes into account the aforementioned dissipation mechanism, which is not accounted for by the standard Mason's model. In our model we add an ideal transformer to the equivalent circuit of the aluminium metallization to represent the coupling between the mechanical mode and the spikes. The winding ratio of the transformer depends on the geometry of the spikes and on the electromechanical coupling factor of the quartz disk. The behavior predicted by our modified model agrees very reasonably with the experimental results obtained for the mechanical quality factor of the charge focusing design.

The model indicates that even when the Al film is in the superconducting state, there is a small ohmic loss between the metalization and the ground, corresponding to a resistance around $\SI{1}{\Omega}$. We associate such losses to resistance in the contact between the bond wires and aluminium layer, or oxidation of the materials. 
Minimization of the aforementioned residual loss would be highly beneficial for cryogenic experiments that use focusing of the mechanical mode to achieve high electromechanical coupling, while the same time need large mechanical quality factors. 
For low temperatures the ohmic losses dominate the dissipation of the mechanical resonator. The increase in temperature increases the quartz intrinsic loss, evening its contribution to the total dissipation with the ohmic losses.

We can conclude that the importance of the electrical dissipation channel is mostly relevant when high-$Q$ piezoelectric resonators are coated with low-resistivity materials and/or patterns that create the opportunity for non-zero low impedance to ground. Metals that at room temperature have enough resistivity to suppress the charge current from the spikes to ground may become a source of losses at cryogenic temperatures, where resistivities are usually lower. The energy leak can be lessened by minimizing the impedance to ground from the coating layer, or  eliminated by using coating films having a high electrical resistivity.


\bigskip

\begin{acknowledgments}

This work was supported by the Academy of Finland (contract 250280, CoE LTQ, 275245), the European Research Council (615755-CAVITYQPD), the Centre for Quantum Engineering at Aalto University, and by the Finnish Cultural Foundation (Central Fund 00160903). We acknowledge funding from the European Union's Horizon 2020 research and innovation program under grant agreement No. 732894 (FETPRO HOT). The work benefited from the facilities at the OtaNano—Micronova Nanofabrication Center and at the Low Temperature Laboratory.

\end{acknowledgments}

\bibliography{sample}

\end{document}